\documentclass[prl,amsmath,amssymb,aps,superscriptaddress,floatfix,reprint,notitlepage]{revtex4-1}

\usepackage[caption=false]{subfig}
\usepackage{ulem} 
\usepackage{amsfonts,amssymb,amsmath}
\usepackage{graphics}
\usepackage{graphicx}
\usepackage{nicefrac}
\usepackage{cases}
\usepackage{mathrsfs}
\usepackage{enumerate}
\usepackage{mathtools}
\usepackage{textcomp}
\usepackage{verbatim}
\usepackage{bm}          
\usepackage{soul}
\usepackage{cancel}
\usepackage{upgreek}
\usepackage{txfonts}
\usepackage{color}
\usepackage[colorlinks=true, urlcolor=blue, linkcolor=blue]{hyperref}
\usepackage[papersize={8.5in,11in}]{geometry}
\usepackage{esint}
\renewcommand{\vec}[1]{\boldsymbol{#1}}
\newcommand {\be} {\begin{equation}}
\newcommand {\ee} {\end{equation}}
\newcommand {\e} {\varepsilon}

\newcommand {\addQ} {\textcolor{red}}

\usepackage{makecell}

\begin{document}

\title{Chiral Wigner crystal phases induced by Berry curvature}

\author{Sandeep Joy}
\affiliation{Department of Physics, Ohio State University, Columbus, OH 43210, USA}
\affiliation{National High Magnetic Field Laboratory, Tallahassee, Florida 32310, USA}
\affiliation{Department of Physics, Florida State University, Tallahassee, Florida 32306, USA}


\author{Leonid Levitov}
\affiliation{Department of Physics, Massachusetts Institute of Technology, Cambridge, MA 02139}

\author{Brian Skinner}
\affiliation{Department of Physics, Ohio State University, Columbus, OH 43210, USA}

\date{\today}
\begin{abstract}

We consider the impact of Berry phase on the Wigner crystal (WC) state of a two-dimensional electron system.
We consider first a model of Bernal bilayer graphene with a perpendicular displacement field, and we show that Berry curvature leads to a new kind of WC state in which the electrons acquire a spontaneous orbital angular momentum when the displacement field exceeds a critical value.  We determine the phase boundary of the WC state in terms of electron density and displacement field at low temperature. 
We then derive the general effective Hamiltonian that governs the ordering of the physical electron spin. We show that this Hamiltonian includes a chiral term that can drive the system into chiral spin-density wave or spin liquid phases. The phenomena we discuss are relevant for the valley-polarized Wigner crystal phases observed in multilayer graphene.
\end{abstract}
\maketitle

\begin{figure}[htb]
\centering
\includegraphics[width=1.0 \columnwidth]{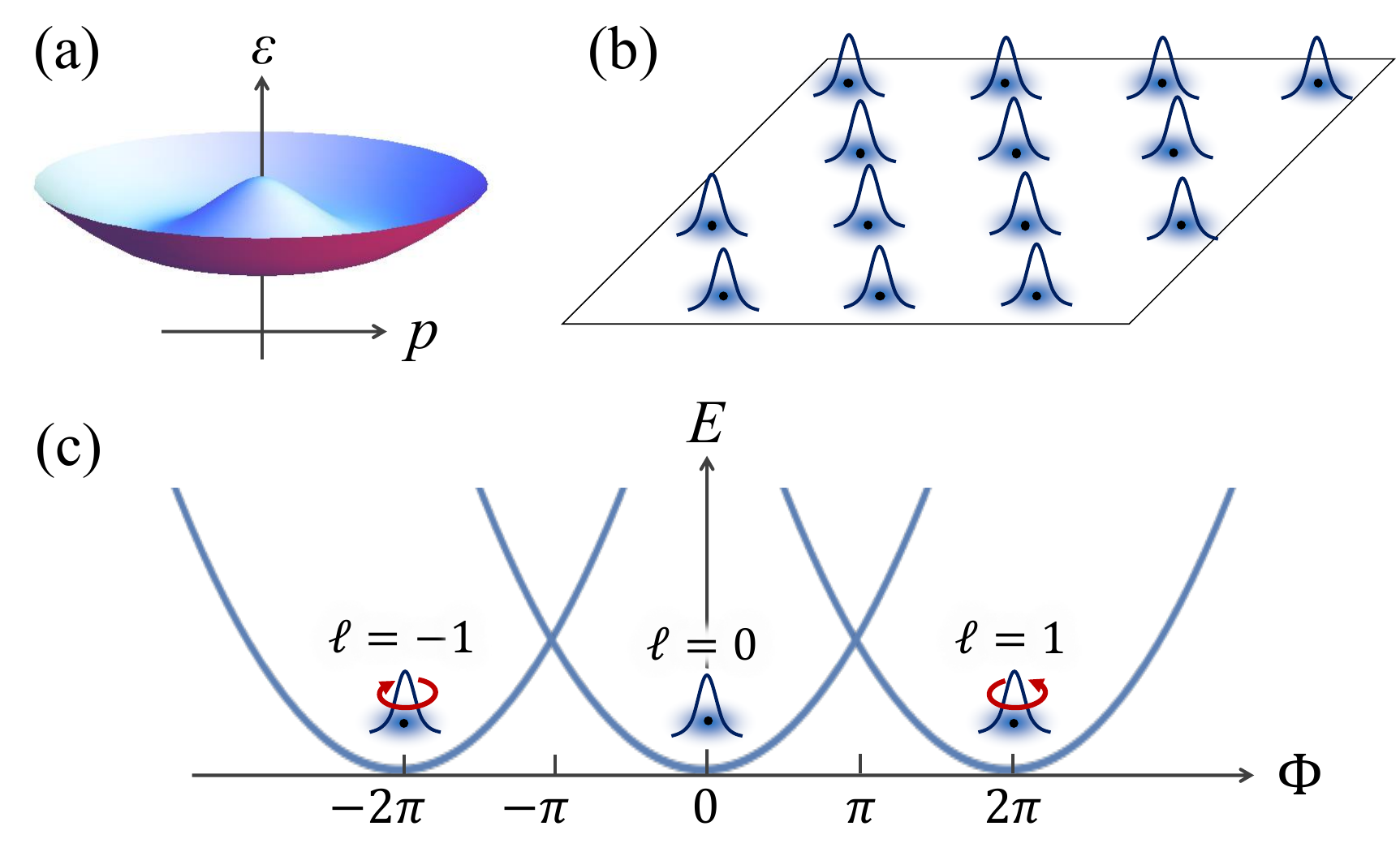}
\caption{(a) Schematic illustration of the conduction band dispersion relation $\varepsilon\left(p\right)$ for electrons in BBG with a perpendicular displacement field [Eq.~(\ref{eq: MHdispersion})]. (b) Schematic depiction of the WC, which can be described as a triangular lattice of nearly-independent quantum harmonic oscillators, each in the ground state of the confining potential created by its neighbors. (c) The ground state energy of an electron in the WC as a function of its angular momentum $\ell$ and the Berry flux $\Phi$ through the interior of its wave function [see Eq.~(\ref{eq:spectrum_wi})]. When $|\Phi| > \pi$, the ground state acquires finite angular momentum, $\ell = \pm 1$.}
\label{fig:Main}
\end{figure}

The Wigner crystal (WC) is perhaps the prototypical strongly correlated electron phase. First proposed in 1934 \cite{Wigner1934On}, the WC arises in situations where the Coulomb interaction between neighboring electrons is much stronger in magnitude than the electrons' kinetic energy. In such situations the electron system minimizes its energy by spontaneously breaking translation symmetry and crystallizing into a regular lattice called the Wigner lattice [depicted in Fig.~\ref{fig:Main}(b)]. For the traditional two-dimensional electron gas with parabolic dispersion and uniform positive background (i.e., the jellium model), the Wigner crystal phase corresponds to the limit of large values of the dimensionless interaction parameter $r_s = [\pi n \hbar^4 / (m^2 e^4)]^{-1/2}$, where $n$ is the two-dimensional electron concentration, $m$ is the electron band mass, and $e^2$ is the squared electron charge divided by the dielectric constant. The value of $r_s$ can be thought of as the ratio between the interaction energy and the kinetic energy (Fermi energy) at low temperature; $r_s$ becomes large when the electron density is low.

In this paper we consider how the WC state is modified when the electron band has a nontrivial Berry curvature. Our motivation arises from the physics of multilayer graphene, where electron bands can be designed to have large density of states that promotes the formation of strongly correlated electron states. Indeed, dramatic cascades of electronic phase phase transitions have been observed experimentally in twisted bilayer graphene \cite{Cao_unconventional_2018, cao_correlated_2018, yankowitz2019tuning, xiaobo2019superconductors, Xie_spectroscopic_2019,sharpe2019emergent, Serlin_Intrinsic_2020, Stepanov_untying_2020, Nuckolls2020strongly, Park_Flavour_2021, choi2021interactiondriven, xie2021fractional, Oh2021evidence},  Bernal bilayer graphene (BBG) \cite{zhou2022isospin, seiler2022quantum, de2022cascade, lin2023spontaneous, zhou2022isospin, zhang_enhanced_2023, holleis2025nematicity, Nam2018afamily, icking2022transport, seiler2023interactiondriven, seiler2024signatures}, and in multilayer rhombohedral graphene 
\cite{shi_electronic_2020, zhou_superconductivity_2021, han_orbital_2023, liu_spontaneous_2024, winterer_ferroelectric_2024, han_large_2024, patterson_superconductivity_2024, han_signatures_2025, xie_tunable_2025, Han2024correlated, Lu2024fractional, Lu2025extended}. These experiments have prompted intense theoretical interest, with most theory works focusing on either the nature of the observed superconducting state (see, e.g., Ref.~\cite{pantaleon_superconductivity_2023} for a recent review) or on the transition between isospin-polarized states within the metallic phase (e.g., \cite{jung_persistent_2015, szabo_competing_2022, Dong2023Isospin, dong_collective_2023}). 


Here we focus on the WC state, and we show that the Berry phase associated with the electron dispersion relation can produce new WC phases with unusual properties. We begin by considering a model of BBG, and we show that when the band gap parameter (controlled in practice by the strength of a perpendicular displacement field) exceeds a critical value, each electron acquires a nonzero orbital angular momentum. The angular momentum is of opposite signs in opposite momentum valleys, e.g., $\ell=+1$ in valley $K$ and $\ell=-1$ in valley $K'$. 
If the graphene band is valley-polarized -- as it often is in both moiré and non-moiré systems \cite{Zhou2021half, Han2023orbital, Arp2024Intervalley, Han2024correlated, Lu2024fractional, Han2025signatures, Lu2025extended, Choi2025Superconductivity} -- all electrons in the WC have the same angular momentum. This causes a sudden change in magnetization at the critical displacement field.

We also show that the Berry phase generically leads to an unusual Hamiltonian for the physical electron spin, which includes new terms that are not found in nonchiral magnetic systems. For valley polarized electrons, this Hamiltonian implies a phase diagram with chiral spin-ordered phases and, potentially, chiral spin liquid phases.

In order to study the phase diagram and orbital ordering of the WC state, let us first consider a model of the dispersion relation of BBG, which consists of two untwisted, A-B stacked graphene layers. BBG (and other multilayer graphenes) are conducive to WC formation because a perpendicular displacement field flattens the bottom of the conduction band and leads to an abnormally small kinetic energy for a given low electron density. Specifically, in BBG the displacement field creates a difference $U$ in potential energy between the top and bottom layers, and the resulting dispersion relation for the conduction band is \cite{McCann2006}
\begin{equation}
\varepsilon\left(p\right)=\left(\frac{\gamma_1^2}{2} + \frac{U^{2}}{4} + v^2p^{2} - \left(\frac{\gamma_1^{4}}{4} +  v^2 p^{2} \left(\gamma_1^2 + U^2\right) \right)^{1/2}\right)^{1/2},
\label{eq:disp_app}
\end{equation}
where $\gamma_{1}$ is the interlayer tunneling amplitude and $v$ is the single-layer graphene Dirac velocity \cite{mccann2013electronic}.
In the remainder of this paper, except where noted explicitly, we use dimensionless units where $\hbar = v = \gamma_1 = 1$ so that all energies are in units of $\gamma_1 \approx 400$\,meV, and all densities are in units of $n_0 \equiv (\gamma_1/\hbar v)^2 \approx 4 \times 10^{13}$\,cm$^{-2}$.
Equation \ref{eq:disp_app} describes a ``Mexican hat'' (MH) shape (see Fig.~\ref{fig:Main}a), with a ring of minima located at a certain value $|\vec{p}| =p_0$. For momenta close to this ring of minima, the dispersion can be expanded as
\begin{equation}
\varepsilon\left(p\right)\simeq \frac{U}{2\sqrt{1+U^{2}}} + \frac{\left(p-p_0\right)^2}{2m},
\label{eq: MHdispersion}
\end{equation}
where $p_0 = (U/2)[(2+U^2)/(1 + U^2)]^{1/2}$ and $m = (1 + U^2)^{3/2}/[2U(2+U^2)]$.
Here and below, we neglect trigonal warping; we comment briefly on the effects of trigonal warping at the end of this paper and leave a detailed discussion to an accompanying work \cite{joy2023wigner}.

The conventional semiclassical model for the WC in two dimensions describes individual electrons as localized to points in a triangular lattice that minimizes the classical electrostatic energy (see Fig.~\ref{fig:Main}b). In this arrangement, each electron resides in a local minimum of the electrostatic potential created by all other electrons. The lowest-order quantum correction to the energy of the WC can be estimated by describing each electron as a harmonic oscillator (HO) residing in a locally parabolic potential whose strength is determined by the Coulomb interaction \cite{mahan1990many, Flambaum1999Spin}.  (In general, one needs to take into account the effects of screening of the Coulomb interaction; we discuss these in the Supplemental Material.) Thus, the lowest-order quantum correction to the energy per electron can be approximated by the ground state energy of a two-dimensional HO. In the conventional WC, the HO description correctly gives the lowest-order quantum correction with an accuracy better than 10$\%$ \cite{mahan1990many}. 
The HO picture also offers a simple method to estimate the critical density associated with quantum melting of the WC. 
Specifically, melting is associated with the Lindemann ratio $\eta = \sqrt{\langle r^2 \rangle}/a$ becoming larger than a critical value $\eta_c$ (the Lindemann criterion), where $\sqrt{\langle r^2 \rangle}$ is the typical radius of the HO ground state and $a = (\sqrt{3} n / 2)^{-1/2}$ is the lattice constant of the Wigner lattice. The value of $\eta_c$ typically falls within the range $0.20 - 0.25$ for any two-dimensional freezing-melting transition \cite{babadi2013universal, AstrakharchikQuantum2007, Joy2022Wigner}.
In this way, our discussion of the WC state is reduced to solving a single-particle problem: that of a single-particle harmonic oscillator in a confining potential created by Coulomb interactions with its neighbors \footnote{It is worth noting that the HO description gives an energy that is equivalent to calculating the Hartree energy of a trial wavefunction that consists of Gaussian wave packets at every site of the Wigner lattice \cite{skinner2013effect, skinner2016interlayer, Maki1983Static, Chitra2005Zero}.}.

When considering an electron with an arbitrary dispersion relation $\e(p)$ in a parabolic confining potential, it is simplest to write the Hamiltonian in momentum space:
\begin{equation}
H = \e(p) + \frac12 k \hat{\vec{r}}^2,
\label{eq:H_eff_1}
\end{equation}
where $\hat{\vec{r}}$ is the position operator and $k$ is the confinement strength. For the WC problem, the value of $k$ is determined by the Coulomb repulsion with neighboring electrons and is generally of order $e^2 n^{3/2}$ -- an exact expression for $k$ is given in the Supplementary Material.
For a band that has nonzero Berry curvature, 
an effective Hamiltonian can be found by writing the position operator $\hat{\vec{r}}$ in momentum space and then projecting the resulting Hamiltonian to the band of interest (see Supplemental Material for details). This procedure yields 
\begin{equation}
H = \varepsilon\left(\vec{p}\right)+\frac{k}{2}\left(\dot{\iota}\vec{\nabla}_{\vec{p}} + \vec{A}\left(\vec{p}\right)\right)^{2},
\label{eq:H_eff_2}
\end{equation}
where $\vec{A}\left(\vec{p}\right)$ is the Berry connection of the band of interest \cite{Karplus1954Hall, Price2014Quantum, Berceanu2016Momentum, Lapa2019Semiclassical}. Comparing Eq.~(\ref{eq:H_eff_2}) to the usual HO Hamiltonian written in position space, one 
observes that the dispersion relation $\e(p)$ acts like a scalar potential in momentum space, while the Berry connection $\vec{A}(\vec{p})$ acts like a magnetic vector potential.
The Berry connection  $\vec{A}$ is particularly straightforward to write in the Coulomb gauge when the Berry curvature $\Omega(\vec{p})$ is radially symmetric (see, for example, Ref.~\cite{knothe2020quartet}) for the expression for the Berry curvature in BBG). In this case
\begin{equation}
\vec{A}=\frac{\Phi\left(p\right)}{2\pi  p}\hat{\phi},
\end{equation}
where $\Phi(p) =\int_{0}^{p}\Omega\left(p'\right) \, 2 \pi p' d\vec{p'}$ is the Berry flux through a disk in momentum space of radius $p$. In BBG, each valley has $\pm 2 \pi$ Berry flux (with opposite signs in opposite valleys), so that $|\Phi(p)|$ is between $0$ and $2 \pi$.

Due to the rotational symmetry of the problem, the corresponding Schr\"{o}dinger equation (SE) can be solved using separation of variables as $\psi(\vec{p})=(g(p)/\sqrt{p}) \exp(\dot{\iota}\ell\phi)$, where $\ell$ is the angular momentum quantum number and $\phi$ is the azimuthal angle in momentum space. 
The resulting SE becomes
\begin{equation}
\left[-\frac{d^{2}}{du^{2}}+\frac{1}{u^{2}}\left(-\frac{1}{4}+\left(-\ell+\frac{\Phi\left(u\sigma\right)}{2\pi}\right)^{2}\right)+\left(u-u_{0}\right)^{2}\right]g(u) = \epsilon g(u).
\label{eq:1DSE_with}
\end{equation}
Here $u$ is a dimensionless momentum ${u \equiv p/\sigma}$, ${u_0 \equiv p_0/\sigma}$, with $\sigma = \sqrt{ m\omega}$ being the characteristic momentum of a HO and $\omega = \sqrt{k/m}$ is the characteristic HO frequency.  We have also defined $\epsilon\equiv E\big/(\hbar \omega/2)$ and we have used Eq.~(\ref{eq: MHdispersion}) to approximate the dispersion relation $\varepsilon(p)$. The full justification of Eq.~(\ref{eq:1DSE_with}) is presented in the Supplemental Material.


As the harmonic confinement experienced by each electron becomes asymptotically weak (i.e., at very low electron density), the value of the constant $u_0$ becomes large, and the eigenstates $g(u)$ are concentrated around $u = u_0$. In other words, the single electron wave function resembles a thin ring in momentum space with radius $p_0$. In this limit Eq.~(\ref{eq:1DSE_with}) becomes precisely the SE for a 1D HO \cite{Chaplik2006bound, Skinner2014Bound, Skinner2019properties}. We can now read off the low-energy spectrum as
\begin{equation}
E_\ell \simeq \frac{\omega}{2}+\frac{\omega}{2u_{0}^{2}}\left(\ell-\frac{\Phi(p_0)}{2\pi}\right)^{2}-\frac{\omega}{8u_{0}^{2}}.
\label{eq:spectrum_wi}
\end{equation}

Equation (\ref{eq:spectrum_wi}) implies that the Berry flux $\Phi$ enclosed by the wavefunction in momentum space plays a crucial role in determining the electron's energy spectrum. In particular, if $|\Phi| > \pi$, the ground state has $|\ell| = 1$ rather than $\ell = 0$, as illustrated in Fig.~\ref{fig:Main}c. Thus, a transition from zero to finite angular momentum can be induced by increasing the displacement field $U$, which increases $p_0$ and thereby results in more Berry flux being enclosed within the wavefunction. Evaluating this condition numerically gives a critical value $U = U_c \approx 1.07$ associated with the transition.

\begin{figure}[htb]
\centering
\includegraphics[width=1.0 \columnwidth]{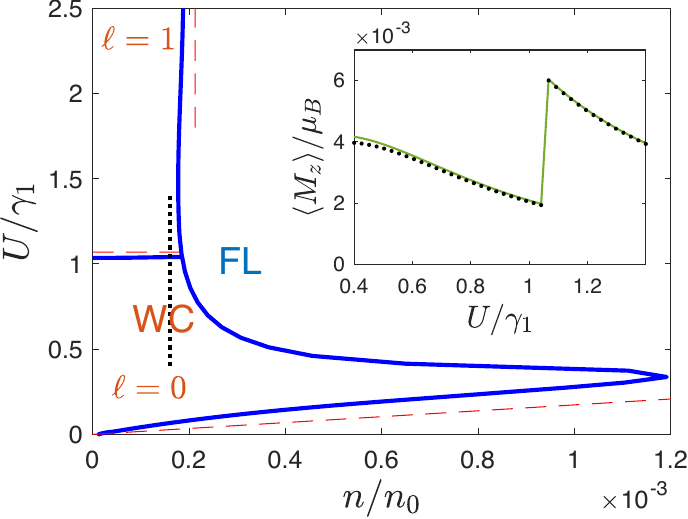}
\caption{The phase diagram of the WC in BBG in the space of electron density $(n)$ and displacement field $(U)$. The blue lines represent the phase boundaries between the $\ell = 0$ WC, $\ell = \pm 1$ WC, and Fermi Liquid (FL) phases, calculated by numerical solution of the Schrodinger equation. The dashed red lines correspond to analytical approximations derived in the Supplemental Material. The inset shows the jump in the magnetization per electron as $U$ passes through the critical value $U_c$, calculated along the line cut indicated by the black dotted line in the main figure. The solid green line corresponds to the analytical result of Eq.~(\ref{eq:mag}), and the dots are results from inserting the numerical solution of the wave function into Eq.~(\ref{eq:mag_full}). 
} 
\label{fig: PD}
\end{figure}

We can map out the full phase diagram for the WC phase by implementing a numerical solution of the effective one-dimensional radial SE [see Eq.~(\ref{eq:1DSE_with}), with $\varepsilon(p)$ given by Eq.~(\ref{eq:disp_app})] for a given density $n$ and interlayer potential $U$.  As mentioned above, the stability of the WC phase is estimated by numerically calculating $\langle r^2 \rangle$ for the ground state wavefunction and using the Lindemann criterion (with $\eta_c = 0.23$ \cite{babadi2013universal}). We ascertain the ground state's angular momentum, whether $\ell = 0$ or $\ell = 1$, by comparing energies obtained numerially. The result is shown in Fig.~\ref{fig: PD}, which generally shows the WC phase occupying a regime of low density and not-too-low displacement field. The extension of the WC state toward large $n$ at small $U$ is associated with the effective mass at the bottom of the band becoming very heavy as $U$ is reduced. The disappearance of the WC state as $U \rightarrow 0$ arises due to interband dielectric screening, which truncates the long-ranged part of the Coulomb interaction when the band gap vanishes (this screening is discussed in more detail in the Supplemental Material).

The discontinuous change in angular momentum at $U = U_c$ leads to observable effects in the magnetization of the WC state. The magnetization operator can be expressed as
\begin{equation}
\hat{M}_z=\frac{e}{2}\left(\vec{v}_{\vec{p}}\times\vec{r}\right),
\end{equation}
where $\vec{v}_{\vec{p}}=\vec{\nabla}_{\vec{p}}\varepsilon(\vec{p})$ is the velocity operator. The expectation value of the magnetization is
\begin{equation}
\left\langle \hat{M}_z\right\rangle =\frac{e}{2}\int \frac{d^2\vec{p}}{\left(2\pi\right)^2}\,v_{p}\left(-\ell+\frac{\Phi(p)}{2\pi} \right) \left|\psi(\vec{p})\right|^{2}.
\label{eq:mag_full}
\end{equation}
Details of the above derivation, along with some analytical results, are provided in the Supplementary Material. The magnetization has two contributions, one from the angular momentum of the wavefunction and the other from the underlying Berry curvature. Tuning the value of $U$ across the $\ell=0$ to $\ell=1$ transition results in a jump in the magnetization. In the limit of $u_0\gg1$ (small electron density or large displacement field), we can evaluate $\langle \hat{M}_{z}\rangle$ analytically as
\begin{equation}
\left\langle \hat{M}_z\right\rangle \simeq\frac{e}{2m}\left(\frac{\sigma^{2}}{2p_{0}^{2}}\right)\left[\ell-\frac{\Phi(p_0)}{2\pi} +\Omega\left(p_{0}\right)p_{0}^{2}\right].
\label{eq:mag}
\end{equation}
Equation (\ref{eq:mag}) implies that at the critical field $U_c$, the magnetization has a jump of magnitude $(e/2m)\left(\sigma^{2}/2p_{0}^{2}\right)$. This jump is depicted in the inset of Fig.~\ref{fig: PD}.

We now consider the impact of the Berry phase on the ordering of the physical electron spin in the WC state. Our considerations in the remainder of this paper make no assumption about the specific nature of the dispersion relation, so that our conclusions apply beyond BBG and are independent of the magnetization transition discussed above. 

\begin{figure}[t!]
\centering
\includegraphics[width=0.95 \columnwidth]{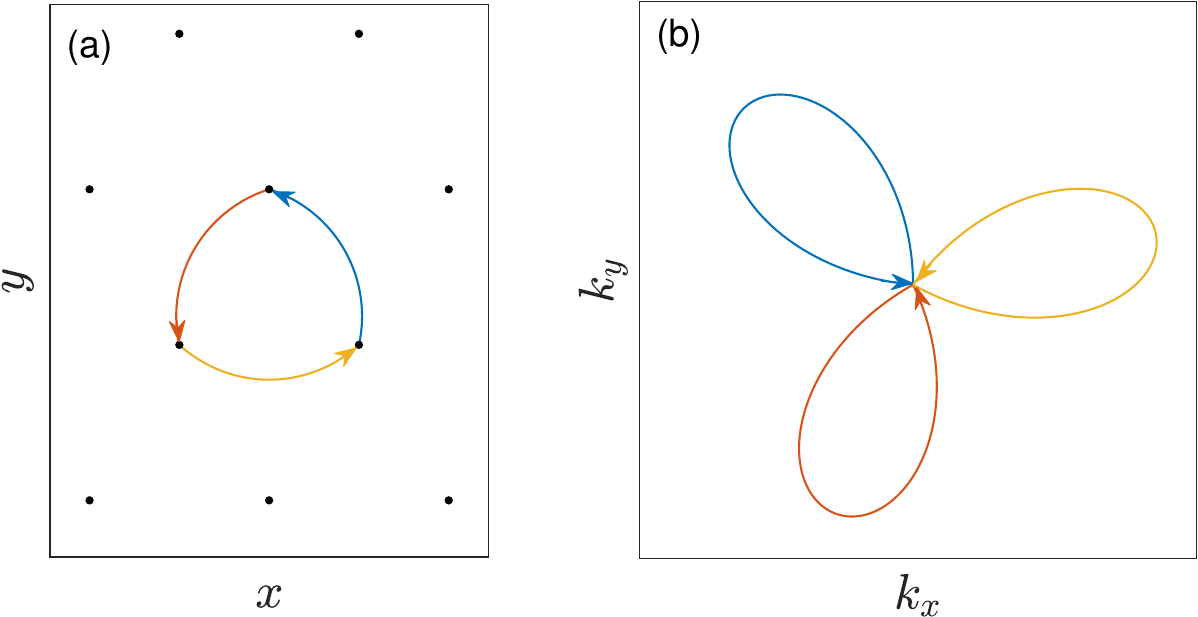}
\caption{Schematic depiction of a three-particle exchange process $J_3$ among neighboring electrons in the WC. Arrows indicate least-action, semiclassical tunneling trajectories. The real-space trajectories are depicted in (a) (with black dots indicating the Wigner lattice positions), while (b) shows the same trajectories in reciprocal space.
} 
\label{fig: tunnelingpaths}
\end{figure}

In the conventional WC, the electron spin is described by the Hamiltonian \cite{Kim2022interstitial, Kim2024dynamical}
\be 
H_\textrm{spin} = \sum_{a}(-1)^{n_a} J_a \left(\mathcal{P}_a + \mathcal{P}_a^{-1} \right),
\ee 
where $a$ labels a ring exchange process involving $n_a$ electrons, $J_a > 0$ is the exchange constant for such processes, and $\mathcal{P}_a$ is the permutation operator associated with the exchange process. 
At large $r_s$ (deep inside the Wigner crystal phase), the dominant exchange processes are a 2-particle exchange between nearest-neighboring electrons and a 3-particle ring exchange among nearest neighbors that form an equilateral triangle (see Fig.\ \ref{fig: tunnelingpaths}a). The Hamiltonian can therefore be approximated as
\be 
H_\textrm{spin} \simeq J_2 \sum_{\langle i, j \rangle} \left( \mathcal{P}_{ij} + \mathcal{P}_{ji} \right) - J_3 \sum_{i,j,k \in \triangle, \triangledown} \left( \mathcal{P}_{ijk} + \mathcal{P}_{kji} \right).
\label{eq:Hspin23}
\ee 
The two-electron permutation operator $\mathcal{P}_{ij} = \mathcal{P}_{ji} = 1/2 + 2 \vec{S}_i \cdot \vec{S}_{j}$, where $\vec{S}_i$ is the spin operator for electron $i$ \cite{roger1983magnetism}. The three-electron permutation operator $\mathcal{P}_{ijk} = \mathcal{P}_{ij}\mathcal{P}_{jk} = 1/4 + \vec{S}_i \cdot \vec{S}_j + \vec{S}_j \cdot \vec{S}_k + \vec{S}_i \cdot \vec{S}_k - 2 i \vec{S}_i \cdot (\vec{S}_j \times \vec{S}_k )$, and $\mathcal{P}_{kji} = \mathcal{P}_{ijk}^\dagger$. Thus, in the conventional WC, the (imaginary) chiral term $\vec{S}_i \cdot (\vec{S}_j \times \vec{S}_k )$ cancels from the spin Hamiltonian, and Eq.~(\ref{eq:Hspin23}) is equivalent to a simple nearest-neighbor Heisenberg Hamiltonian, 
$H_\textrm{spin} \simeq 2 (J_2 - J_3) \sum_{\langle i, j \rangle} \vec{S}_i \cdot \vec{S}_j + \text{const.}$

At not-to-low electron density (only moderately large $r_s$), the value of $J_2$ is apparently larger than $J_3$, so that the ground state ordering is antiferromagnetic (AFM). However, at asymptotically large $r_s$, the order is inverted, $J_3 \gg J_2$, leading to ferromagnetic (FM) ordering. This dominance of ring exchange over direct exchange at large $r_s$ is a unique feature of the WC and the long-range Coulomb interaction that produces it. It arises because the semiclassical trajectory associated with three neighboring electrons simultaneously exchanging their positions has a lower tunneling action than that of the two-electron exchange \cite{Kim2022interstitial} (e.g., rotating three electrons in the confining potential of their triangular-lattice neighbors is ``almost free''). Evidence for this transition from AFM to FM order of the WC as a function of increasing $r_s$ has been seen in quantum Monte Carlo calculations \cite{drummond2009phase}, and recent experiments have reported evidence of AFM order near the transition \cite{Zhang2025transport, falson_competing_2022} and FM order at larger $r_s$ \cite{falson_competing_2022, Hossain2020observation}.

However, if the electron system experiences a consistent sign of Berry curvature, as arises when the electrons are valley polarized, there is explicit time reversal symmetry breaking that introduces a relative phase $2\phi$ between the clockwise and counterclockwise three-particle exchange. Consequently, the spin Hamiltonian becomes
\footnote{Similar chiral terms can appear in the effective spin Hamiltonian when there is a magnetic field that produces an Aharonov-Bohm phase, for example in the Hubbard model with a magnetic field \cite{olexei2006orbital}. In this context such terms are suppressed by a factor $t/U$ compared to direct exchange, where $t$ is the nearest-neighbor hopping and $U$ is the on-site energy.}
\begin{align}
H_\textrm{spin}  = & \, J_2 \sum_{\langle i, j \rangle} \left( \mathcal{P}_{ij} + \mathcal{P}_{ji} \right) - J_3 \sum_{i,j,k \in \triangle, \triangledown} \left( e^{i \phi} \mathcal{P}_{ijk} + e^{-i \phi} \mathcal{P}_{kji} \right) \nonumber \\
= & \, 2 \left(J_2 - J_3 \cos \phi \right) \sum_{\langle i,j \rangle} \vec{S}_i \cdot \vec{S}_j  \nonumber \\  
& + 4 J_3 \sin\phi\ \sum_{i,j,k \in \triangle, \triangledown} \vec{S}_i \cdot \left( \vec{S}_j \times \vec{S}_k \right) \nonumber \\
\equiv & \, J \sum_{\langle i,j \rangle} \vec{S}_i \cdot \vec{S}_j  + J_\chi \sum_{i,j,k \in \triangle, \triangledown} \vec{S}_i \cdot \left( \vec{S}_j \times \vec{S}_k \right).
\label{eq:Hspinchiral}
\end{align}
Thus, the Berry phase introduces a chiral term in the spin Hamiltonian, which can lead to chiral ordering or chiral spin textures, as has been pointed out in the Fermi liquid setting \cite{Chiral_Stoner_PhysRevB.110.104420,panigrahi2024spinchiralityfermionstirring}.
The nearest-neighbor exchange constant $J=2(J_2 -J_3\cos{\phi})$ can be either positive or negative, depending on the value of $r_s$ (i.e., on the electron density), and the chiral term $J_\chi = 4 J_3 \sin \phi$ has a sign given by the sign of the Berry curvature (and therefore on the choice of valley $K$ or $K'$). The magnitude of the Berry phase $\phi$ can be estimated from the area subtended by the reciprocal-space tunneling paths (depicted in Fig.\ \ref{fig: tunnelingpaths}b) for three-particle exchange. This estimate gives $\phi \propto 1/r_s$, as we discuss in the Supplementary Material. 

For positive $J$ (AFM nearest-neighbor coupling), the Hamiltonian of Eq.~(\ref{eq:Hspinchiral}) has been studied previously by density matrix renormalization group techniques \cite{KunYang_PhysRevB.96.075116, kuhlenkamp2024chiral}, which found a wide window of parameters $0.3 \lesssim J_\chi / J \lesssim 0.6$ for which the ground state is a chiral spin liquid phase \cite{KunYang_PhysRevB.96.075116}. 
To our knowledge, the Hamiltonian of Eq.~(\ref{eq:Hspinchiral}) has not been studied for FM (negative) $J$. However, we expect that for $J < 0$, increasing $|J_\chi/J|$ is associated with a transition from a FM state to a state with a nontrivial chiral spin density wave pattern. The nature of this transition, and whether a spin liquid phase arises in some critical window of $|J_\chi/J|$, remains to be seen.


In closing, let us comment on the experimental implications of the predictions made here. Our discussion so far has focused on the WC state at zero temperature, for which the melting of the WC with increasing density arises from quantum fluctuations. At finite temperatures, thermal fluctuations can also melt the WC state. At densities not too close to the critical density associated with quantum melting (at a given value of $U$), one can estimate the melting temperature by setting the Lindemann ratio to be $\eta=\sqrt{\langle r^{2}\rangle _{\text{thermal}}}/a$, where the amplitude of classical fluctuations is estimated using the equipartition theorem: $k\langle r^{2}\rangle _{\text{thermal}} = k_B T$. The maximum melting temperature can be estimated by using the largest density $n_c$ of the WC state (here, $n_c \approx 0.0012$, see Fig.~\ref{fig: PD}). This procedure gives a maximum melting temperature on the order of $\sim 10$\,K, with the melting temperature decreasing proportional to $n^{1/2}$ as the density is reduced. Experimentally, the WC state can be inferred by a combination of distinctive transport and thermodynamic measurements, including negative compressibility Refs.~\cite{bello_density_1981, kravchenko1990evidence, eisenstein1992negative, eisenstein1994compressibility, shapira1996thermodynamics, Skinner2010anomalously, Li2011very, skinner2013giant} and sharp ``pinning'' behavior in the $I$-$V$ curve \cite{yoon_wigner_1999, knighton_evidence_2018, falson_competing_2022}.  The experiments of Ref.~\cite{zhou2022isospin} observed negative compressibility in BBG emerging at low temperature and high displacement field but not coexisting with an insulating temperature dependence. Ref.~\cite{seiler2022quantum} reports a state with insulating-like temperature dependence that emerges within a window of relatively high displacement field and low densities, which the authors characterize as being consistent with a WC. Evidence for WC formation has also been reported in rhombohedral 4- and 5-layer graphene \cite{Han2024correlated, Lu2024fractional, Lu2025extended}, which may be an ideal platform for realizing the physics we discuss here, given their relatively flat dispersion and increased Berry phase at the Dirac point.

Throughout this paper we have neglected the effects of trigonal warping on the WC state. In general, trigonal warping splits the  rotationally-symmetric ``Mexican hat" band structure into three discrete mini-valleys, so that the orbital angular momentum $\ell$ is no longer a good quantum number at sufficiently low energy. However, even in the presence of trigonal warping, an electron in a radially symmetric confining potential can still undergo a magnetization transition with increasing displacement field associated with winding of the wave function in momentum space. This magnetized state competes with alternative WC states having nontrivial mini-valley ordering \cite{calvera2022pseudo}. We explore this competition, and the effects of trigonal warping more generally, in an accompanying paper \cite{joy2023wigner}.

We emphasize, however, that the chiral term appearing in the spin Hamiltonian [Eq.~(\ref{eq:Hspinchiral})] does not rely on rotational symmetry of the band structure or on the existence of an $\ell \neq 0$ ground state; it appears generically when the electrons are valley polarized. The magnitude $J_\chi$ of the chiral term is generally small in the Wigner crystal phase, since it is proportional to $1/r_s \ll 1$. However, the effects of this term may still be significant, since the competing direct exchange $J$ term apparently vanishes at a particular value of $r_s \approx 38$ \cite{drummond2009phase}, which is not far from the onset of the Wigner crystal phase $r_s \approx 32$. It is therefore plausible that spin liquid physics is already arising in the spin order of valley-polarized Wigner crystal phases that have been realized in multilayer graphenes \cite{Han2024correlated, Lu2024fractional, Lu2025extended}. 

\acknowledgments
\textit{Acknowledgments.} \;  The authors are grateful to Zachariah Addison, Debanjan Chowdhury, Zhiyu Dong, Liang Fu, Aaron Hui, Long Ju, Kyle Kawagoe, Steve Kivelson, Austin Merkel, Aatmaj Rajesh, T.\ Senthil, and Jairo Velasco for useful discussions. B.S. acknowledges support from the US National Science Foundation under grant no.\ DMR-2045742. S.J. acknowledges support from Florida State University through the Quantum Postdoctoral Fellowship, and from the National High Magnetic Field Laboratory supported by the US National Science Foundation (grant no.\ DMR-2128556) and the State of Florida.
\bibliography{WC+OM}

\end{document}


\widetext
\begin{center}
\textbf{\large Supplementary Material for ``Chiral Wigner crystal phases induced by Berry curvature"}
\end{center}
\setcounter{equation}{0}
\setcounter{figure}{0}
\setcounter{table}{0}
\setcounter{page}{1}
\makeatletter
\renewcommand{\theequation}{S\arabic{equation}}
\renewcommand{\thefigure}{S\arabic{figure}}
\renewcommand{\bibnumfmt}[1]{[S#1]}
\renewcommand{\citenumfont}[1]{S#1}

\begin{center}

Sandeep Joy$^{1,2,3}$, Leonid Levitov$^{4}$, Brian Skinner$^{1}$ \\

\textit{$^{1}$Department of Physics, Ohio State University, Columbus, OH 43210, USA} \\
\textit{$^{2}$National High Magnetic Field Laboratory, Tallahassee, Florida 32310, USA}\\
\textit{$^{3}$Department of Physics, Florida State University, Tallahassee, Florida 32306, USA}\\
\textit{$^{4}$Department of Physics, Massachusetts Institute of Technology, Cambridge, MA 02139}


(Dated: \today)
\end{center}

\section{S1. Derivation of the effective harmonic oscillator Hamiltonian in a band with Berry curvature}
\label{sec:HO_BC}

In this section, we derive the effective harmonic oscillator (HO) Hamiltonian in momentum space when the electrons originate from a band with non-zero Berry curvature \cite{Karplus1954Hall, Price2014Quantum, Berceanu2016Momentum, Lapa2019Semiclassical}. The single particle Hamiltonian is given by $H=H_{0}+U(r)$, where $H_{0}$ comes from the underlying crystal lattice and $U(r)=kr^{2}/2$ is the harmonic confinement potential. The Bloch eigenstates of $H_{0}$ are denoted by $|n,p\rangle$  ($n$ and $\vec{p}$ denote the band index and quasi-momentum), which satisfies $H_{0}\exp(\dot{\iota}\vec{p}\cdot\vec{r})|n,p\rangle =E_{n}(\vec{p})\exp(\dot{\iota}\vec{p}\cdot\vec{r})|n,p\rangle$, where $E_{n}(\vec{p})$ is the band dispersion of the $n^{\text{th}}$ band. Any eigenstates of $H$ can be expanded in the eigenbasis of $H_{0}(\vec{p})$ as
\begin{equation}
\left|\psi\right\rangle =\sum_{n}\sum_{\vec{p}}\psi_{n}\left(\vec{p}\right)\exp\left(\dot{\iota}\vec{p}\cdot\vec{r}\right)\left|n,p\right\rangle, 
\end{equation}
where $\psi_{n,\vec{p}}$ denotes the expansion coefficients. The eigenvalue equation for the Hamiltonian $H$ can be written as
\begin{equation}
\sum_{n}\sum_{\vec{p}}E_{n}\left(\vec{p}\right)\psi_{n}\left(\vec{p}\right)\exp\left(\dot{\iota}\vec{p}\cdot\vec{r}\right)\left|n,p\right\rangle +U\left(r\right)\sum_{n}\sum_{\vec{p}}\psi_{n}\left(\vec{p}\right)\exp\left(\dot{\iota}\vec{p}\cdot\vec{r}\right)\left|n,p\right\rangle =\epsilon\sum_{n}\sum_{\vec{p}}\psi_{n}\left(\vec{p}\right)\exp\left(\dot{\iota}\vec{p}\cdot\vec{r}\right)\left|n,p\right\rangle. 
\end{equation}
For brevity, let us denote $|\chi_{n,\vec{p}}\rangle =\exp(\dot{\iota}\vec{p}\cdot\vec{r})|n,p\rangle$. Taking an inner product with $\langle \chi_{m,\vec{q}}|$, the above equation yields
\begin{equation}
E_{m}\left(\vec{q}\right)\psi_{m}\left(\vec{q}\right)+\sum_{n}\sum_{\vec{p}}\left\langle \chi_{m,\vec{q}}\right|U\left(r\right)\left|\chi_{n,\vec{p}}\right\rangle \psi_{n}\left(\vec{p}\right)=\epsilon\psi_{m}\left(\vec{q}\right).
\end{equation}
We used the orthonormality condition of the Bloch eigenstates $\langle \chi_{m,\vec{q}}|\chi_{n,\vec{p}}\rangle =\delta(\vec{p}-\vec{q})\delta_{m,n}$. We are left with the evaluation of $\langle \chi_{m,\vec{q}}|U(r)|\chi_{n,\vec{p}}\rangle$. Let us first evaluate $\langle \chi_{m,\vec{q}}|\widehat{\vec{r}}|\chi_{n,\vec{p}}\rangle$. Using the identity $\int d\vec{r}|\vec{r}\rangle \langle \vec{r}|=\mathbb{I}$ and $\langle \vec{r}|\chi_{n,\vec{p}}\rangle =\exp(\dot{\iota}\vec{p}\cdot\vec{r})w_{n,\vec{p}}(\vec{r})$, where $w_{n,\vec{p}}(\vec{r})\equiv\langle \vec{r}|n,\vec{p}\rangle$ we obtain
\begin{equation}
\left\langle \chi_{m,\vec{q}}\right|\widehat{\vec{r}}\left|\chi_{n,\vec{p}}\right\rangle=\int d\vec{r}\exp\left(-\dot{\iota}\vec{q}\cdot\vec{r}\right)w_{m,\vec{q}}^{*}\left(\vec{r}\right)\vec{r}\exp\left(\dot{\iota}\vec{p}\cdot\vec{r}\right)w_{n,\vec{p}}\left(\vec{r}\right).
\end{equation}
Using the following relation, $\vec{r}\exp(\dot{\iota}\vec{p}\cdot\vec{r})=-\dot{\iota}\vec{\nabla}_{\vec{p}}\exp(\dot{\iota}\vec{p}\cdot\vec{r})$, the integrand in the previous equation can be rewritten
\begin{equation}
\left\langle \chi_{m,\vec{q}}\right|\widehat{\vec{r}}\left|\chi_{n,\vec{p}}\right\rangle=\dot{\iota}\vec{\nabla}_{\vec{p}}\left(\int d\vec{r}\exp\left(\dot{\iota}\left(\vec{p}-\vec{q}\right)\cdot\vec{r}\right)w_{m,\vec{q}}^{*}\left(\vec{r}\right)w_{n,\vec{p}}\left(\vec{r}\right)\right)+\dot{\iota}\int d\vec{r}\exp\left(\dot{\iota}\left(\vec{p}-\vec{q}\right)\cdot\vec{r}\right)w_{m,\vec{q}}^{*}\vec{\nabla}_{\vec{p}}w_{n,\vec{p}}\left(\vec{r}\right).
\label{eq:A5}
\end{equation}
The first term in the above equation can be evaluated to be
\begin{equation}
-\dot{\iota}\vec{\nabla}_{\vec{p}}\left(\int d\vec{r}\exp\left(\dot{\iota}\left(\vec{p}-\vec{q}\right)\cdot\vec{r}\right)w_{m,\vec{q}}^{*}\left(\vec{r}\right)w_{n,\vec{p}}\left(\vec{r}\right)\right)=\dot{\iota}\delta\left(\vec{q}-\vec{p}\right)\vec{\nabla}_{\vec{p}}\delta_{m,n}.
\end{equation}
The second integral in Eq.~\ref{eq:A5} vanishes unless $\vec{p}=\vec{q}$, because $w_{n,\vec{p}}(\vec{r})$ is a periodic function, which gives
\begin{equation}
\dot{\iota}\int d\vec{r}\exp\left(\dot{\iota}\left(\vec{p}-\vec{q}\right)\cdot\vec{r}\right)w_{m,\vec{q}}^{*}\vec{\nabla}_{\vec{p}}w_{n,\vec{p}}\left(\vec{r}\right)=\dot{\iota}\delta\left(\vec{q}-\vec{p}\right)\left\langle m,\vec{q}\right|\vec{\nabla}_{\vec{q}}\left|n,\vec{q}\right\rangle.
\end{equation}
Recognizing the (non-abelian) Berry connection $A_{m,n}(\vec{q})=\dot{\iota}\langle m,\vec{q}|\vec{\nabla}_{\vec{q}}|n,\vec{q}\rangle$ \cite{Xiao2010Berry}, we arrive at
\begin{equation}
\left\langle \chi_{m,\vec{q}}\right|\widehat{\vec{r}}\left|\chi_{n,\vec{p}}\right\rangle =\delta\left(\vec{q}-\vec{p}\right)\left(\dot{\iota}\delta_{m,n}\vec{\nabla}_{\vec{p}}+A_{m,n}\left(\vec{p}\right)\right).
\end{equation}
Similarly, inserting the completeness identity twice,  $\widehat{\vec{r}}^2$ can be shown to be
\begin{equation}
\sum_{n}\sum_{\vec{p}}\left\langle \chi_{m,\vec{q}}\right|\widehat{\vec{r}}^2\left|\chi_{n,\vec{p}}\right\rangle \psi_{n}\left(\vec{p}\right)=\sum_{n,n'}\left(\dot{\iota}\delta_{m,n'}\vec{\nabla}_{\vec{q}}+A_{m,n'}\left(\vec{q}\right)\right)\cdot\left(\dot{\iota}\delta_{n',n}\vec{\nabla}_{\vec{q}}+A_{n',n}\left(\vec{q}\right)\right)\psi_{n}\left(\vec{q}\right),
\end{equation}
leading to the following SE
\begin{equation}
E_{m}\left(\vec{q}\right)\psi_{m}\left(\vec{q}\right)+\frac{k}{2}\sum_{n,n'}\left(\dot{\iota}\delta_{m,n'}\vec{\nabla}_{\vec{q}}+A_{m,n'}\left(\vec{q}\right)\right)\cdot\left(\dot{\iota}\delta_{n',n}\vec{\nabla}_{\vec{q}}+A_{n',n}\left(\vec{q}\right)\right)\psi_{n}\left(\vec{q}\right)=\epsilon\psi_{m}\left(\vec{q}\right).
\end{equation}
So far, we have not made any assumptions in deriving the above equation. Now, let's assume that the relevant band index is $m$, and in the second term, only $\psi_{m}(\vec{q})$ is non-negligible. This lead to
\begin{equation}
E_{m}\left(\vec{q}\right)\psi_{m}\left(\vec{q}\right)+\frac{k}{2}\sum_{n'}\left(\dot{\iota}\delta_{m,n'}\vec{\nabla}_{\vec{q}}+A_{m,n'}\left(\vec{q}\right)\right)\cdot\left(\dot{\iota}\delta_{n',m}\vec{\nabla}_{\vec{q}}+A_{n',m}\left(\vec{q}\right)\right)\psi_{m}\left(\vec{q}\right)\approx\epsilon\psi_{m}\left(\vec{q}\right).
\end{equation}
Separating $n'=m$ and $n'\neq m$ terms, we can write
\begin{align}
    \begin{split}
        E_{m}\left(\vec{q}\right)\psi_{m}\left(\vec{q}\right)+\frac{k}{2}\left(\dot{\iota}\vec{\nabla}_{\vec{q}}+A_{m,m}\left(\vec{q}\right)\right)^{2}\psi_{m}\left(\vec{q}\right)+\frac{k}{2}\sum_{n'\neq m}\left(A_{m,n'}\left(\vec{q}\right)\right)\cdot\left(A_{n',m}\left(\vec{q}\right)\right)\psi_{m}\left(\vec{q}\right)&\approx\epsilon\psi_{m}\left(\vec{q}\right),\\\implies\left[E_{m}\left(\vec{q}\right)+\frac{k}{2}\left(\dot{\iota}\vec{\nabla}_{\vec{q}}+A_{m,m}\left(\vec{q}\right)\right)^{2}+\frac{k}{2}\sum_{n'\neq m}\left|A_{m,n'}\left(\vec{q}\right)\right|^{2}\right]\psi_{m}\left(\vec{q}\right)&\approx\epsilon\psi_{m}\left(\vec{q}\right).
    \end{split}
\end{align}
We have finally arrived at the effective Hamiltonian
\begin{equation}
H_{eff}=E_{m}\left(\vec{q}\right)+\frac{k}{2}\left(\dot{\iota}\vec{\nabla}_{\vec{q}}+A_{m,m}\left(\vec{q}\right)\right)^{2}+\frac{k}{2}\sum_{n'\neq m}\left|A_{m,n'}\left(\vec{q}\right)\right|^{2}.
\end{equation}
The energy dispersion and Berry connection of the $m^{th}$ band act as a momentum space scalar and magnetic vector potential. There is another contribution to the momentum space scalar potential given by
\begin{equation}
E_{add}\left(\vec{q}\right)=\frac{k}{2}\sum_{n'\neq m}\left|A_{m,n'}\left(\vec{q}\right)\right|^{2}.
\end{equation}
This additional term is the same as the trace of the underlying quantum geometric tensor \cite{Classen2015Position}. Using the following representation of $A_{n,n'}$, we can recast the additional term into a gauge invariant form
\begin{equation}
A_{n',n}=\dot{\iota}\frac{\left\langle n',\vec{p}\right|\left(\vec{\nabla}_{\vec{p}}H_{0}\left(\vec{p}\right)\right)\left|n,p\right\rangle }{\left(E_{n}\left(\vec{p}\right)-E_{n'}\left(\vec{p}\right)\right)}.
\end{equation}
The additional potential can be rewritten as
\begin{align}
    \begin{split}
        E_{add}\left(\vec{q}\right)&=\frac{k}{2}\sum_{n'\neq m}\left|A_{m,n'}\left(\vec{q}\right)\right|^{2},\\&=\frac{k}{2}\sum_{n'\neq m}\frac{\left|\left\langle m,\vec{q}\right|\left(\vec{\nabla}_{\vec{p}}H_{0}\left(\vec{p}\right)\right)\left|n',q\right\rangle \right|^{2}}{\left(E_{n'}\left(\vec{q}\right)-E_{m}\left(\vec{q}\right)\right)^{2}}.
    \end{split}
\end{align}
This last term $E_{add}(\vec{q})$ can be neglected in the WC regime where the electron density $n$ is small, since $E_{add}$ scales linearly with $k \propto n^{3/2}$ while the terms we are focuse on are proportional to $\omega\propto \sqrt{k}\propto n^{3/4}$.

\section{S2. Schr\"{o}dinger Equation for the Harmonic Oscillator with Mexican hat dispersion}
\label{sec: MHHO}

In this section, we solve the problem of a particle confined in a Harmonic oscillator potential with a Mexican hat (MH) dispersion (the case in the presence of Berry curvature is presented in the next section). To that end, let us consider the following Hamiltonian:
\begin{equation}
\hat{H}=\frac{1}{2m}\left(\left|\hat{\vec{p}}\right|-p_{0}\right)^{2}+\frac{1}{2}m\omega^{2}\hat{\vec{r}}^{2},
\end{equation}
where $\hat{\vec{p}}$ and $\hat{\vec{r}}$ are the momentum and position operators, $\omega$ is the angular frequency, and $m$ is the mass of the particle. Since the position operator assumes the form of a gradient operator in momentum space, $\hat{\vec{r}}=\dot{\iota}\hbar\vec{\nabla}{\vec{p}}$, this problem look like a particle in a MH potential in momentum space. The corresponding Schrödinger equation (SE) in momentum space is given by:
\begin{equation}
\left[-\frac{1}{2}\hbar^{2}m\omega^{2}\left(\frac{\partial^{2}}{\partial p^{2}}+\frac{1}{p}\frac{\partial}{\partial p}+\frac{1}{p^{2}}\frac{\partial^{2}}{\partial\phi^{2}}\right)+\frac{1}{2m}\left(p-p_{0}\right)^{2}\right]\psi\left(\vec{p}\right)=E\psi\left(\vec{p}\right).
\label{eq: SE_MS}
\end{equation}
Due to the rotational symmetry of the problem, the solution to the above SE is of the variable separable form, $\psi(\vec{p})=f(p)e^{\dot{\iota}\ell\phi}$, and the allowed values of $\ell$ are integers. In the first step, we turn the above SE into an effective one-dimensional SE. In order to do that, we use the ansatz $f=g/\sqrt{p}$, which allows us to eliminate the first-order derivative term. Substituting this ansatz, the SE becomes:
\begin{equation}
-\frac{1}{2}\hbar^{2}m\omega^{2}\left[\frac{d^{2}g}{dp^{2}}+\frac{1}{p^{2}}\left(\frac{1}{4}-\ell^2\right)g\right] + \frac{1}{2m}\left(p-p_{0}\right)^{2}g=Eg.
\end{equation}
We can recast the effective radial (in momentum space) SE in a dimensionless form by dividing all the momentum scales by $\sigma\equiv\sqrt{\hbar m\omega}$, so that $u\equiv p/\sigma$ and $u_{0}\equiv p_{0}/\sigma$. We obtain:
\begin{equation}
-\left[\frac{d^{2}g}{du^{2}}+\frac{1}{u^{2}}\left(\frac{1}{4}-\ell^{2}\right)g\right]+\left(u-u_{0}\right)^{2}g=\epsilon g
\label{Eq: SESimple},
\end{equation}
where we have defined $\epsilon\equiv E\big/(\hbar\omega/2)$. In the limit where the MH is deep enough, $u_{0}>>1$ (the wavefunction is a thin ring in momentum space), we can find the eigenstates and eigenenergy perturbatively in $1/u_0$. Let us expand the above equation around $u_0$ by taking $x\equiv u-u_{0}$ and keeping terms up to $1/u_0^2$:
\begin{equation}
-\frac{d^{2}g}{dx^{2}}+x^{2}g=\tilde{\epsilon}g,
\label{eq:1D_HO}
\end{equation}
where we defined:
\begin{equation}
\tilde{\epsilon}=\epsilon+\frac{1}{4u_{0}^{2}}-\frac{\ell^{2}}{u_{0}^{2}}.
\end{equation}
The SE in Eq.~\ref{eq:1D_HO} is the equation for the one-dimensional Harmonic oscillator, and its eigenvalues are given by:
\begin{equation}
\tilde{\epsilon} = 2n+1, \quad n\in\mathbb{Z}^{+}.
\end{equation}
Going back to the original units, the energy eigenvalues are given by:
\begin{equation}
E_{n,\ell}=\frac{\hbar\omega}{2}\left(2n+1\right)+\frac{\hbar\omega\ell^{2}}{2u_{0}^{2}}-\frac{\hbar\omega}{8u_{0}^{2}}.
\end{equation}
This result has a nice physical interpretation. When $p\sim p_{0}$, the particle effectively experiences harmonic confinement in the radial direction and behaves dispersionless in the azimuthal direction. Therefore, the ground state energy of this system corresponds to a one-dimensional harmonic oscillator. The corresponding normalized wavefunction for $n=0$ principal quantum number is given by:
\begin{equation}
\psi_\ell\left(p,\phi\right)=\left(\frac{2\sqrt{\pi}}{\sigma}\right)^{1/2}\frac{\exp\left[-\frac{\left(p-p_{0}\right)^{2}}{2\sigma^{2}}\right]}{\sqrt{p}}\:\exp\left[\dot{\iota}\ell\phi\right].
\label{eq:psi_p}
\end{equation}
By Fourier transformation, we can calculate the real space wavefunction:
\begin{align}
    \begin{split}
     \tilde{\psi}_\ell\left(r,\theta\right)&=\frac{1}{\left(2\pi\right)^{2}}\int d^{2}\vec{p} \, e^{\dot{\iota}\vec{p}.\vec{r}}\psi\left(\vec{p}\right),\\&=\left(\frac{2\sqrt{\pi}}{\sigma}\right)^{1/2}\frac{1}{\left(2\pi\right)^{2}}\int_{0}^{2\pi}d\phi\int_{0}^{\infty}pdp\frac{\exp\left[-\frac{\left(p-p_{0}\right)^{2}}{2\sigma^{2}}\right]}{\sqrt{p}}\:e^{\dot{\iota}\left(pr\cos\left(\theta-\phi\right)+\ell\phi\right)}.
    \end{split}
\end{align}
The angular integral can be simplified as follows:
\begin{equation}
    \int_{0}^{2\pi}d\phi\:\exp\left[\dot{\iota}\left(pr\cos\left(\phi-\theta\right)+\ell\phi\right)\right]=\exp\left[\dot{\iota}\ell\theta\right]\int_{0}^{2\pi}d\tilde{\phi}\:\exp\left[\dot{\iota}\left(pr\cos\left(\tilde{\phi}\right)+\ell\tilde{\phi}\right)\right].
\end{equation}
We use the following Jacobi–Anger expansion \cite{arfken1999mathematical} to perform the above angular integration:
\begin{equation}
\exp\left[\dot{\iota}a\cos x\right]=\sum_{n=-\infty}^{\infty}\dot{\iota}^{n}J_{n}\left(a\right)\exp\left[\dot{\iota}nx\right],
\end{equation}
where $J_{n}$ is the $n^{th}$ Bessel function of the first kind. The result of the angular integral is given by:
\begin{equation}
\int_{0}^{2\pi}d\tilde{\phi}\:\exp\left[\dot{\iota}\left(pr\cos\left(\tilde{\phi}\right)+\ell\tilde{\phi}\right)\right]=2\pi\dot{\iota}^{-\ell}\left(-1\right)^{\ell}J_{\ell}\left(pr\right).
\end{equation}
Ignoring the phase factors, we have:
\begin{equation}
 \tilde{\psi}_\ell\left(r,\theta\right)=\exp\left[\dot{\iota}\ell\theta\right]\left(\frac{2\sqrt{\pi}}{\sigma}\right)^{1/2}\frac{1}{2\pi}\int_{0}^{\infty}dp\sqrt{p}\exp\left[-\frac{\left(p-p_{0}\right)^{2}}{2\sigma^{2}}\right]J_{\ell}\left(pr\right).
\end{equation}
To perform this integral, we again introduce the dimensionless variables $x$ and $u_0$ ($u_0\equiv p_0/\sigma$ and $x\equiv (p-p_0)/\sigma$) that we used before:
\begin{equation}
\int_{0}^{\infty}dp\sqrt{p}\exp\left[-\frac{\left(p-p_{0}\right)^{2}}{2\sigma^{2}}\right]J_{\ell}\left(pr\right)=\sigma^{3/2}\int_{-u_{0}}^{\infty}dx\sqrt{u_{0}+x}\exp\left[-\frac{x^{2}}{2}\right]J_{\ell}\left(\left(u_{0}+x\right)\sigma r\right).
\end{equation}
For large $x$, $J_{\ell}\left(x\right)$ has the following asymptotic form:
\begin{equation}
J_{\ell}\left(x\right)\approx\sqrt{\frac{2}{\pi x}}\cos\left[x-\frac{\ell\pi}{2}-\frac{\pi}{4}\right].
\end{equation}
This approximation is justified as long as $p_{0}r\gg1$. Using the asymptotic expansion, we have:
\begin{equation}
\int_{-u_{0}}^{\infty}dx\sqrt{u_{0}+x}\exp\left[-\frac{x^{2}}{2}\right]J_{\ell}\left(\left(u_{0}+x\right)\sigma r\right)\approx\sqrt{\frac{2}{\pi\sigma r}}\cos\left[u_{0}\sigma r-\frac{\ell\pi}{2}-\frac{\pi}{4}\right]\int_{-\infty}^{\infty}dx\exp\left[-\frac{x^{2}}{2}\right]\cos\left[x\sigma r\right].
\end{equation}
Using
\begin{equation}
\int_{-\infty}^{\infty}dx\cos\left[ax\right]e^{-\frac{x^{2}}{2}}=\sqrt{2\pi}\exp\left[-\frac{a^{2}}{2}\right],
\end{equation}
we arrive at the final expression:
\begin{equation}
\tilde{\psi}_\ell\left(r,\theta\right) = \left(\frac{p_{0}\sigma}{\sqrt{\pi}}\right)^{1/2}J_{\ell}\left(p_{0}r\right)\exp\left[-\frac{\sigma^{2}r^{2}}{2}\right]\exp\left[\dot{\iota}\ell\theta\right],
\label{eq: real_psi_1}
\end{equation}
or equivalently
\begin{equation}
\tilde{\psi}_\ell\left(r,\theta\right)=\left(\frac{2\sigma}{\pi\sqrt{\pi}}\right)^{1/2}\frac{\cos\left[p_{0}r-\frac{\ell\pi}{2}-\frac{\pi}{4}\right]}{\sqrt{r}}\exp\left[-\frac{\sigma^{2}r^{2}}{2}\right]\exp\left[\dot{\iota}\ell\theta\right].
\label{eq: real_psi_2}
\end{equation}

The width of the wavefunction in real space can be calculated as
\begin{equation}
\left\langle \hat{r}^{2}\right\rangle_{\ell} \simeq\frac{1}{\sigma ^2}\left(\frac{1}{2}+\frac{\ell^{2}}{u_{0}^{2}}-\frac{1}{4u_{0}^{2}}\right).
\label{eq: nclargeU}
\end{equation}
We can use this result, together with the Lindemann criterion, to estimate the critical density associated with the WC phase. 
In the limit of large $U \gg 1$, where the mass in the radial direction approaches $m \simeq 1/2$ and dielectric screening is unimportant, this calculation yields 
\be 
n_c \approx 2.1\times10^{-4}.
\ee 
This value of $n$ is indicated by a vertical dashed red line in Fig.~2 of the main text.

\subsection{Schr\"{o}dinger Equation for the Harmonic Oscillator with Mexican hat dispersion and Berry curvature}

In this subsection, we derive the eigenenergies of an electron with a Mexican hat dispersion that is subject to a harmonic confining potential in the presence of Berry curvature. Following the previous section, in the momentum space, $\hat{\vec{r}}^2$ should be replaced by $(\dot{\iota}\vec{\nabla}_{\vec{q}}+A(\vec{q})^{2}$. Exploiting the rotational symmetry of the system, in the Coulomb gauge, the Berry connection can be written as
\begin{equation}
\vec{A}=\frac{\Phi\left(p\right)}{2\pi p}\hat{\phi},
\end{equation}
where
\begin{equation}
\Phi\left(p\right)=\int_{0}^{p}dp'\,2\pi p'\,\Omega\left(p'\right),
\end{equation}
is the amount of Berry flux through a disk in momentum space of radius $p$. Modifying the SE to incorporate the effect of Berry curvature amount to rewriting the azimuthal part of the gradient operator in the following way
\begin{equation}
\frac{\dot{\iota}}{p}\frac{\partial}{\partial\phi_{p}}\longrightarrow\frac{\dot{\iota}}{p}\frac{\partial}{\partial\phi_{p}}+\frac{\Phi\left(p\right)}{2\pi p}.
\end{equation}
The effective one-dimensional harmonic oscillator SE in this situation can be shown to be  (following the same procedure as in Sec~\ref{sec: MHHO})
\begin{equation}
\left[-\frac{d^{2}}{du^{2}}+\frac{1}{u^{2}}\left(-\frac{1}{4}+\left(-\ell+\frac{\Phi\left(u\sigma\right)}{2\pi }\right)^{2}\right)+\left(u-u_{0}\right)^{2}\right]g=\epsilon g.
\label{eq:1DSE_BC}
\end{equation}
Let us again expand around $u_0$, changing the variable, $x\equiv u-u_{0}$
\begin{equation}
-\left[\frac{d^{2}g}{dx^{2}}+\frac{1}{u_{0}^{2}}\left(\frac{1}{4}-\left(-\ell+\frac{\Phi\left(p_{0}\right)}{2\pi}\right)^{2}\right)g\right]+x^{2}g=\epsilon g.
\end{equation}
With the following definition
\begin{equation}
\tilde{\epsilon}_\ell=\epsilon+\frac{1}{4u_{0}^{2}}-\frac{\left(-\ell+\frac{\Phi\left(p_0\right)}{2\pi}\right)^{2}}{u_{0}^{2}},
\end{equation}
we get the following familiar differential equation
\begin{equation}
-\frac{d^{2}g}{dx^{2}}+x^{2}g=\tilde{\epsilon}_{\ell}g.
\end{equation}
We once again encounter the 1D simple harmonic oscillator SE, wherein eigenvalues are given by
\be
\tilde{\epsilon}_{n,l}=\left(2n+1\right).
\ee

The critical field associated with the transition to finite angular momentum can be found by the condition: 
\begin{equation}
\Phi(p_0) = \int_{0}^{p_{0}\left(U_{c}\right)}dp' 2 \pi p' \Omega\left(p'\right) = \pi.
\end{equation}
(In the hypothetical case where the Berry curvature is constant as a function of $p$, this condition simplifies to $\Omega\left(p_0\right) p_0^2 =1$.) Evaluating this condition numerically gives
\be 
U_c \approx 1.07.
\label{eq: Uc}
\ee 
This value of $U$ is indicated by a horizontal dashed red line in Fig.~2 of the main text.

\section{S5. Orbital magnetization an electron wavepacket inside a Wigner crystal}
\label{sec: OrbMag}

\begin{figure}[htb]
\centering
\includegraphics[width=0.6 \columnwidth]{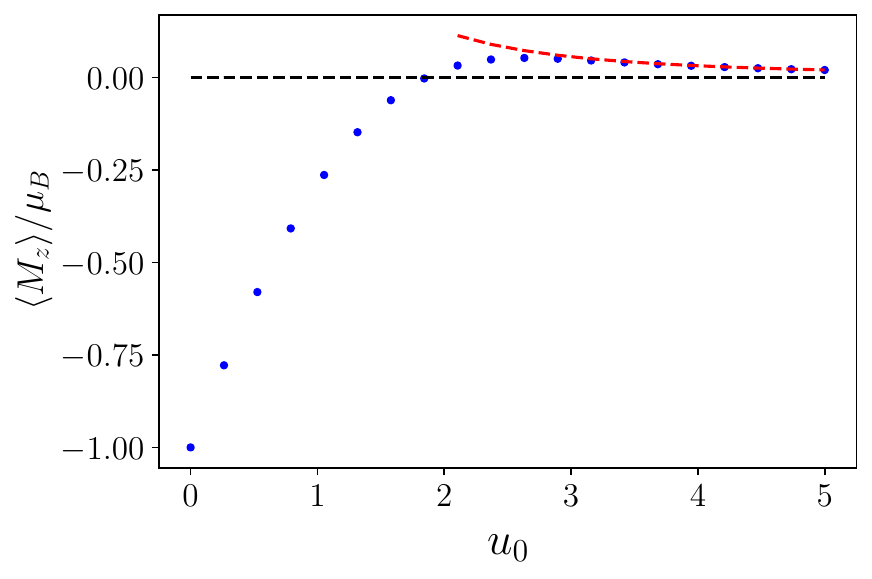}
\caption{Average magnetization $\langle\hat{M}_z\rangle$ of the first excited state ($\ell=1$) of an electron with Mexican hat dispersion as a function of the dimensionless Mexican hat radius $u_0$. Blue dots show results from a numeric calculation. The limit $u_0=0$ corresponds to the usual parabolic dispersion and gives the usual value $\langle\hat{M}_z\rangle=\mu_{B}$, resembling the case of a conventional 2D HO. As the radius increases, the magnetization decreases in magnitude and undergoes a sign change around $u_0 \simeq 1$. At larger values of $u_0$, $\langle\hat{M}_z\rangle$ converges towards the analytical prediction of Eq.~\ref{eq: mag_ana} (represented by a dashed red line).} 
\label{fig: Sign_change}
\end{figure}

Here we discuss the orbital magnetization of the WC state, focusing on the Mexican hat dispersion. We show how the sign of the magnetization is inverted with increasing $p_0$. Orbital magnetization is the negative of the partial derivative of free energy with respect to an applied magnetic field (which is a linear response to an external magnetic field).
\begin{equation}
\hat{M}_z=\left.-\frac{\partial F}{\partial B}\right|_{B=0}.
\end{equation}
At zero temperature and using minimal substitution, we can calculate the free energy as
\begin{align}
    \begin{split}
        F&=H\left(\vec{p}+e\vec{A}\right),\\&\simeq H\left(\vec{p}\right)+\frac{e}{2}\left(\vec{A}\cdot\vec{\nabla}_{\vec{p}}H+\left(\vec{\nabla}_{\vec{p}}H\right)\cdot\vec{A}\right).
    \end{split}
\end{align}
where $\vec{\nabla}_{\vec{p}}H=\vec{v}_{p}$ is the velocity operator and $-e$ is the charge of the particle. For a  constant magnetic field $\vec{B}=B\hat{z}$, and choosing $\vec{A}=-\frac{1}{2}(\vec{r}\times\vec{B})$ we can derive
\begin{equation}
F\simeq H\left(\vec{p}\right)+\frac{e\vec{B}\cdot}{2}\left(\left(\frac{1}{2}\left(\vec{r}\times\vec{v}_{p}\right)\right)-\left(\frac{1}{2}\left(\vec{v}_{p}\times\vec{r}\right)\right)\right).
\end{equation}
Using the fact that $(\vec{r}\times\vec{v}_{p})_{z}=-(\vec{v}_{p}\times\vec{r})_{z}$, the magnetization operator can be found to be 
\begin{equation}
\hat{M}_z=\frac{e}{2}\left(\vec{v}_{p}\times\vec{r}\right)\hat{z}.
\end{equation}
For rotationally symmetric system $v_{\vec{p}}=v_{p}\hat{p}$, where $v_{p}=\frac{\partial E}{\partial p}$ and in the presence of Berry curvature, $m$ can be written 
\begin{equation}
\hat{M}_z = \frac{e}{2}\left(\frac{v_{p}}{p}\left(\dot{\iota}\frac{\partial}{\partial\phi}+\frac{\Phi\left(p\right)}{2\pi}\right)\right).
\end{equation}
When $\psi(\vec{p})=f(p)e^{\dot{\iota}\ell\phi}$, we get
\begin{equation}
\left\langle \hat{M}_z\right\rangle =\frac{e}{2}\,\frac{1}{\left(2\pi\right)^2}\int_{0}^{2\pi}d\phi\int_{0}^{\infty}dp\,v_{p}\left(-\ell+\frac{\Phi\left(p\right)}{2\pi}\right)\left|f\left(p\right)\right|^{2}.
\end{equation}
Plugging the wavefunction from Eq.~\ref{eq:psi_p}, and expanding $\Phi(p)\simeq\Phi(p_{0})+2\pi(p-p_{0})p_{0}\Omega(p_{0})$, one arrives at
\begin{equation}
\left\langle \hat{M}_z\right\rangle =\frac{e}{2m}\left(\frac{1}{\sqrt{\pi}\sigma}\right)\int_{0}^{\infty}dp\,\left(\frac{p-p_{0}}{p}\right)\exp\left[-\frac{\left(p-p_{0}\right)^{2}}{\sigma^{2}}\right]\left(-\ell+\frac{\Phi\left(p_{0}\right)}{2\pi}+\left(p-p_{0}\right)p_{0}\Omega\left(p_{0}\right)\right).
\end{equation}
Thus we finally arrive at
\begin{equation}
\left\langle \hat{M}_z\right\rangle = \frac{e}{2m}\left(\frac{\sigma^{2}}{2p_{0}^{2}}\right)\left(\ell-\frac{\Phi\left(p_{0}\right)}{2\pi}+\Omega\left(p_{0}\right)p_{0}^{2}\right).
\label{eq: mag_ana}
\end{equation}
Even in the absence of Berry curvature, the aforementioned findings indicate an intriguing transition as the radius of the Mexican hat, denoted as $u_0$, is varied. In the case of small $u_0$, the sign of the magnetization aligns with the physical notion of a negatively charged particle circulating counterclockwise around a loop. However, as $u_0$ increases, the magnitude of the magnetization diminishes and undergoes a sign reversal beyond a critical value. This phenomenon is depicted in Figure ~\ref{fig: Sign_change}. 

\section{S6. Confinement strength and Comment on the dielectric function}
\label{sec:dielectric}

The value of $k$ (confinement strength) is determined by the interactions between electrons in the Wigner lattice and is, therefore, a function of the electron density, defined as follows. If one takes the origin $\vec{r} = 0$ to be a site of the Wigner lattice, then the potential $u(\vec{r})$ experienced by the electron near the origin is $u(\vec{r}) = \sum_{i\neq0}V(|\vec{r} - \vec{R}_{i}|)$, where $i$ indexes the sites of the Wigner lattice, $V(r)$ is the Coulomb interaction, $\vec{R}_i$ denotes a lattice vector of the Wigner lattice, and $\vec{r}$ is the position of the electron near the origin. Here we have assumed that $|\vec{r}|$ is much smaller than the Wigner lattice constant so that other electrons in the Wigner lattice can be treated as having a fixed position. The value of $k$ can then be found by Taylor expanding $u(\vec{r})$ to the second order in $r$, which gives \cite{Joy2022Wigner}
\begin{equation}
k = \frac{1}{2}\sum_{i\neq0}\left[V''\left(\left|\vec{R}_{i}\right|\right)+\frac{V'\left(\left|\vec{R}_{i}\right|\right)}{\left|\vec{R}_{i}\right|}\right].
\label{eq: ho}
\end{equation}
In general, $V(r)$ should be taken as the screened Coulomb interaction, which we discuss below.

When the interlayer potential $U$ is small, the gap between conduction and valence bands is also small, so the electron-electron interaction is significantly modified by screening by virtual interband excitations. This effect is captured by the static polarization function $\Pi(q)$, such that the Fourier-transformed Coulomb interaction is \cite{mahan1990many}
\be 
V(q) = \frac{V_0(q)}{1 - \Pi(q) V_0(q)},
\ee 
where $V_0(q) = 2 \pi e^2/(\epsilon_r q)$ is the unscreened Coulomb interaction. For gapped BBG, $\Pi(q)$ has the following asymptotic behaviors \cite{Silvestrov2017Wigner}: 

\begin{equation}
\Pi\left(q\right)=\left\{ \begin{array}{cc}
-q/4, & q\gg1\\
-\ln(4)/\pi, & 1\gg q\gg\sqrt{U}\\
-4q^{2}/3\pi U, & q \ll \sqrt{U}
\end{array}\right. .
\label{eq: Pi}
\end{equation}
As we show below, the WC state melts at densities $n \ll U$, so that the relevant momentum scale $q \sim n^{1/2}$ corresponds to the final regime in Eq.~\ref{eq: Pi}. Thus, we can use 
the screened interaction:
\begin{equation}
V\left(q\right)=\frac{2\pi e^{2}}{\epsilon_{r}(q+ q^{2}/q_{0} )},
\end{equation}
where $q_{0} = 3U/(8e^2)$. 

Notice that at sufficiently large $q$ (low electron density), the Coulomb potential $V(q) \simeq 3U/(8\epsilon_r q^2)$ corresponds to a logarithmic dependence of $V(r)$ on the spatial distance $r$. This modified Coulomb interaction significantly modifies the critical density associated with WC melting \cite{Joy2022Wigner} whenever $n \gg U^2$. In the limit $U \ll 1$ and $U \ll q \ll \sqrt{U}$, the relevant electron dispersion is quartic, given by $\varepsilon(p)\approx U/2 + p^4/U$ \cite{Silvestrov2017Wigner}. Using the screened interaction for $V(q)$, and calculating the confining potential strength via Eq.~\ref{eq: ho}, one can show that $k=3\pi U n /8$ \cite{Joy2022Wigner}. The Lindemann criterion for melting then gives a critical density
\begin{equation}
n_c = C_2 \times U,
\label{eq: ncSmallU}
\end{equation}
where $C_2\approx 173$. Thus, unlike in the conventional WC problem, in BBG, the critical density vanishes in the limit $U \rightarrow 0$. That is, interlayer dielectric response precludes the formation of a WC state unless a displacement field is applied that provides an energy gap for virtual electron-hole pairs. Equation \ref{eq: ncSmallU} is plotted as a dashed red line in Fig.~2 of the main text.

\section{S7. Semiclassical estimate of the Berry phase for three-electron ring exchange}

In the main text we mention that clockwise and counterclockwise ring exchange processes have a relative phase $2 \phi$ arising from the Berry curvature. Here we estimate the value of $\phi$ for three-electron ring exchange. For simplicity we consider the usual parabolic dispersion relation, $E(k) = \hbar^2 k^2/2m$. 

At large $r_s$, the dominant contribution to the amplitude of the exchange process is associated with the semiclassical path of least action for the exchange process. If we parameterize this path by a parameter $s$, then the (imaginary) momentum $k(s)$ along the path is given by 
\be 
\frac{\hbar^2 k(s)^2}{2 m} = E - V(s),
\ee 
where $E$ is the electron energy and $V(s)$ is the potential energy along the path. The maximum potential energy of the electron along the tunneling trajectory corresponds to a saddle point in the potential energy landscape created by the Coulomb interaction with other electrons in the Wigner lattice. This maximum, relative to the starting energy $E$ of the electron, is generically of order $e^2/R$, where $R \sim n^{-1/2}$ is the Wigner lattice constant. So the maximum value of the imaginary momentum is
\be 
\left| k_\textrm{max} \right| \sim \sqrt{ \frac{m e^2 n^{1/2}}{\hbar^2} }.
\ee 
In (imaginary) reciprocal space, the tunneling path for a given electron corresponds to clockwise or counterclockwise excursion from $\vec{k} = 0$ to a point at distance $k_\textrm{max}$ from the origin and back again, as depicted in Fig.~3b of the main text. 
The Berry phase $\phi$ associated with this trajectory is given by the flux of Berry curvature through all three such trajectories. If we consider the Berry curvature $\Omega$ to have a constant value over a region of size $k_\textrm{max}$ near the origin, then $\phi \sim \Omega \cdot |k_\textrm{max}|^2$, or
\be 
\phi \sim \frac{\Omega e^2 n^{1/2}}{\hbar^2} \sim \frac{\Omega}{a_B^2} \frac{1}{r_s},
\ee 
where $a_B = \hbar^2/(me^2)$ is the effective Bohr radius. As an example, monolayer graphene with sublattice symmetry breaking (e.g., a mass term) has \cite{xiao2007valley} $\Omega/a_B^2 = \alpha^2$, where $\alpha = e^2/(\hbar v)$ is the effective fine structure constant of graphene (with $v$ the graphene velocity), which is generally an order-1 number.

We note that there is, in general, also a Berry phase modification of the two-electron exchange amplitude. At large $r_s$, exchange of two electrons is associated with two least-action exchange trajectories: one clockwise and one counterclockwise. In the presence of Berry curvature these two paths acquire a relative Berry phase $2 \phi_2$. The corresponding term in the spin Hamiltonian then becomes
\be 
H_\textrm{spin}  =  J_2 \sum_{\langle i, j \rangle} \left( \mathcal{P}_{ij} + \mathcal{P}_{ji} \right) 
\Rightarrow J_2 \sum_{\langle i, j \rangle} \left( e^{i \phi_2} \mathcal{P}_{ij} + e^{-i \phi_2} \mathcal{P}_{ji} \right).
\ee
Since for a two-electron exchange $\mathcal{P}_{ij} = \mathcal{P}_{ji}$, the Berry phase $\phi_2$ has only the effect of renormalizing the value of $J_2$:
\be 
J_2 \Rightarrow J_2 \cos(\phi_2).
\ee

\bibliography{WC+OM}